\documentclass[preprintnumbers, prd, floatfix,onecolumn, preprintnumbers, letterpaper,
superscriptaddress,nofootinbib]{revtex4}
\usepackage[format=hang,justification=raggedright,singlelinecheck=false,skip=0cm]{caption}
\usepackage{amsfonts}
\usepackage{amsmath}
\usepackage{amssymb,epsf}
\usepackage{latexsym}
\usepackage{graphicx,epsfig}
\usepackage{amssymb}
\usepackage{float}
\usepackage{subfigure}
\usepackage[colorlinks=true,citecolor=blue,linkcolor=blue,urlcolor=black]{hyperref}

\graphicspath{{Images/}}

\begin{document}

\title{Charged scalar quasi-normal modes for linearly charged
dilaton-Lifshitz solutions}
\author{M. Kord Zangeneh}
\email{mkzangeneh@shirazu.ac.ir}
\affiliation{Physics Department, Faculty of Science, Shahid Chamran University of Ahvaz,
Ahvaz 61357-43135, Iran}
\affiliation{Research Institute for Astronomy and Astrophysics of Maragha
(RIAAM)-Maragha, IRAN, P. O. Box: 55134-441}
\affiliation{Physics Department and Biruni Observatory, Shiraz University, Shiraz 71454,
Iran}
\affiliation{Center of Astronomy and Astrophysics, Department of Physics and Astronomy,
Shanghai Jiao Tong University, Shanghai 200240, China}
\author{B. Wang}
\email{wang\_b@sjtu.edu.cn}
\affiliation{Center of Astronomy and Astrophysics, Department of Physics and Astronomy,
Shanghai Jiao Tong University, Shanghai 200240, China}
\affiliation{Center for Gravitation and Cosmology, College of Physical Science and
Technology, Yangzhou University, Yangzhou 225009, China}
\author{A. Sheykhi}
\email{asheykhi@shirazu.ac.ir}
\affiliation{Physics Department and Biruni Observatory, Shiraz University, Shiraz 71454,
Iran}
\affiliation{Research Institute for Astronomy and Astrophysics of Maragha
(RIAAM)-Maragha, IRAN, P. O. Box: 55134-441}
\author{Z. Y. Tang}
\email{tangziyu@sjtu.edu.cn}
\affiliation{Center of Astronomy and Astrophysics, Department of Physics and Astronomy,
Shanghai Jiao Tong University, Shanghai 200240, China}

\begin{abstract}
Most available studies of quasi-normal modes for Lifshitz black solutions
are limited to the neutral scalar perturbations. In this letter, we
investigate the wave dynamics of massive charged scalar perturbation in the
background of $(3+1)$-dimensional charged dilaton Lifshitz black
branes/holes. We disclose the dependence of the quasi-normal modes on the
model parameters, such as the Lifshitz exponent $z$, the mass and charge of
the scalar perturbation field and the charge of the Lifshitz configuration.
In contrast with neutral perturbations, we observe the possibility to
destroy the original Lifshitz background near the extreme value of charge
where the temperature is low. We find out that when the Lifshitz exponent
deviates more from unity, it is more difficult to break the stability of the
configuration. We also study the behavior of the real part of the
quasi-normal frequencies. Unlike the neutral scalar perturbation around
uncharged black branes where an overdamping was observed to start at $z=2$
and independent of the value of scalar mass, our observation discloses that
the overdamping starting point is no longer at $z=2$ and depends on the mass
of scalar field for charged Lifshitz black branes. For charged scalar
perturbations, fixing $m_s$, the asymptotic value of $\omega_R$ for high $z$
is more away from zero when the charge of scalar perturbation $q_s$
increases. There does not appear the overdamping.
\end{abstract}

\maketitle

\section{Introduction}

In black hole physics, quasi normal mode (QNM) is a powerful tool to study
the evolution of perturbations in the exterior of black holes \cite%
{qnm4,qnm5,qnm6,wang1}. The behavior of QNM can be used to identify the
black hole existence and disclose dynamical stability of black hole
configurations. Besides, QNM can serve as a testing ground of fundamental
physics. It is widely believed that QNM can give deeper understandings of
the AdS/CFT \cite{wang1,3,30,31,32}, dS/CFT \cite{4} correspondences, loop
quantum gravity \cite{5} and also phase transitions of black holes \cite{6}
etc.

In this letter we will examine the QNM of the linearly charged
dilaton-Lifshitz black brane solutions and try to disclose deep influences
of the model parameters on the perturbation wave dynamics and examine the
stability of the background configurations. Asymptotic Lifshitz black
solutions are interesting duals to many condensed matter systems \cite{Lif}.
They are duals to systems with Schrodinger-like scaling symmetries i.e. $%
t\rightarrow \lambda ^{z}t$, $\vec{\mathbf{x}}\rightarrow \lambda \vec{%
\mathbf{x}}$, where $z$ is dynamical critical exponent. Lifshitz spacetime
is not a vacuum solution of Einstein gravity with or without cosmological
constant and some Lifshitz supporting fields are needed. Different Lifshitz
supporting fields have been considered in literatures, such as including
higher curvature corrections \cite{hcc1,hcc2,hcc3}, inserting massive \cite%
{massg1} and massless \cite{tarrio,Kord,Kord1,Kord2,Kord3,Kord4} Abelian
gauge fields coupled to dilaton and non-Abelian $SU(2)$ Yang-Mills fields
coupled to dilaton \cite{nonAb} etc.

The behavior of neutral scalar perturbations has been extensively studied
for Lifshitz solutions in the presence of different Lifshitz supporting
fields. In \cite{nmg1,nmg2}, scalar and spinorial perturbations around ($2+1$%
)-dimensional Lifshitz black holes with $z=3$ in the context of New Massive
Gravity (which includes higher curvature terms) have been explored and it
has been shown that black holes are stable under both of these
perturbations. Moreover, it has been shown that higher-dimensional Lifshitz
black branes are stable under massive scalar perturbations in the presence
of higher curvature corrections \cite{hcc4,hcc5}. In \cite{hcc6}, ($3+1$%
)-dimensional Lifshitz black holes with $z=0$ and in the presence of higher
curvature corrections have been considered. In the case of massive scalar
perturbations minimally coupled to curvature, these black holes are
unstable, whereas such black holes with massless perturbations conformally
coupled to curvature are stable. In the presence of a massive gauge field
including Proca term, scalar massive QNMs for uncharged Lifshitz black
branes with $z=2$ have been studied in \cite{pro}. It was shown that the
QNMs are purely damping (the real part of QNM vanishes) and Lifshitz black
branes are always stable. It was further shown that quasi-normal frequencies
of scalar massive perturbation around topological Lifshitz black holes with $%
z=2$ \cite{Mann} are always purely imaginary and negative, supporting that
these black holes are always stable \cite{topqnm}. In \cite{myung}, scalar
QNMs of $2$- and $3$-dimensional uncharged Lifshitz black branes with $z=3$
in the context of New Massive Gravity and $4$-dimensional uncharged Lifshitz
black branes with $z=2$ in the presence of massive and massless gauge fields
coupled to dilaton have been studied. It has been shown that quasi-normal
frequencies are purely negative imaginary, reflecting that these solutions
are globally stable. Considering dilaton field and massless gauge fields
coupled to dilaton as Lifshitz supporting matter fields, massive QNMs with
zero momenta for higher-dimensional Lifshitz black branes have been examined 
\cite{LifAIM1} and it has been shown that black branes are stable under
these perturbations. In this context, massive and massless QNMs for
higher-dimensional Abelian \cite{LifAIM2} and non-Abelian \cite{LifAIM4}
charged Lifshitz black branes with hyperscaling violation have been studied
as well. They were also found stable under scalar perturbations. Massive and
massless scalar QNMs around ($3+1$)-dimensional dilaton-Lifshitz black
holes/branes in the presence of nonlinear power-law Maxwell field have been
explored in \cite{LifAIM3}. It was shown that these Lifshitz solutions are
always stable. It was further found that QNMs for black branes can become
overdamping when the dynamical critical exponent $z$ and angular momentum
take some special values. In \cite{park}, retarded Green functions of the
current and momentum operator of a Lifshitz field theory have been
investigated and it was shown that there exists a massive QNMs with an
effective mass linearly proportional to temperature in non-vanishing
momentum case.

Most available studies of QNMs for Lifshitz black solutions are limited to
the neutral scalar perturbation. Some extensions to the Fermionic and
electromagnetic perturbations for Lifshitz solutions have been reported
recently \cite{ferm,elec}. It is of great interest to generalize the
discussion to the charged scalar perturbation in the background of Lifshitz
solutions. The charged scalar perturbation has been studied extensively in
other contexts \cite{konop,1405.4931,konop1,1002.2679,1010.2806,1111.6729}.
There has been examples showing that charged scalar perturbation can result
in the spacetime instabilities in the Reissner-Nordstr\"{o}m-de Sitter
configuration and the anti-de Sitter charged black holes in \cite%
{1405.4931,konop1} and \cite{1002.2679,1010.2806,1111.6729}, respectively.
In the AdS space, the charged scalar field can condensate onto the AdS black
hole to form a new hairy black hole. In the present letter, we will examine
the dynamics of massive charged scalar perturbation in the charged Lifshitz
black solution backgrounds. We will disclose the dependence of the QNMs on
the model parameters, such as the mass and charge of the scalar perturbation
field, the charge of the background spacetime etc. Especially we will
examine in the Lifshitz configuration, whether the instability of the
background configuration can appear and how the instability will depend on
model parameters including the Lifshitz exponent $z$. For simplicity in our
discussion, we will focus on the $(3+1)$-dimensional Lifshitz spacetime and
leave the study on the higher dimensional influence on QNMs for our future
work.

The letter is organized as follows. In the next section, we will review
solutions of the four-dimensional linearly charged Lifshitz configurations.
Then we will write out the wave equations for the charged scalar
perturbation and explain the method of the numerical computation we are
going to employ. In Section \ref{nr}, we will report results of QNMs. In the
last section we will conclude our results.

\section{Review on $4$-dimensional linearly charged Lifshitz solutions}

In this section we will review the $4$-dimensional charged dilaton-Lifshitz
solutions. The line element of $4$-dimensional Lifshitz black solutions can
be written as \cite{tarrio}%
\begin{equation}
ds^{2}=-\frac{r^{2z}}{l^{2z}}f(r)dt^{2}+{\frac{l^{2}}{r^{2}}}\frac{dr^{2}}{%
f(r)}+r^{2}d\Omega _{k}^{2},  \label{lifmet}
\end{equation}%
where $z$ is dynamical critical exponent and 
\begin{equation}
d\Omega _{k}^{2}=\left\{ 
\begin{tabular}{ll}
$d\theta ^{2}+\sin ^{2}(\theta )d\phi ^{2}$ & $k=1$ \\ 
$d\theta ^{2}+d\phi ^{2}$ & $k=0$ \\ 
$d\theta ^{2}+\sinh ^{2}(\theta )d\phi ^{2}$ & $k=-1$%
\end{tabular}%
\right. ,
\end{equation}%
represents a $2$-dimensional hypersurface with constant curvature $2k$. $%
f(r) $ in the line element (\ref{lifmet}) has a solution in the context of
Einstein-dilaton gravity in the presence of linear Maxwell electrodynamics
and two Lifshitz supporting gauge fields%
\begin{equation}
S=-\frac{1}{16\pi }\int_{\mathcal{M}}d^{4}x\sqrt{-g}\left[ \mathcal{R}%
-2(\nabla \Phi )^{2}-2\Lambda +-e^{-2\lambda _{1}\Phi
}F-\sum\limits_{i=2}^{3}e^{-2\lambda _{i}\Phi }H_{i}\right] ,  \label{action}
\end{equation}%
in which $\mathcal{R}$ is the Ricci scalar on manifold $\mathcal{M}$, $\Phi $
is the dilaton field and $F$ and $H_{i}$'s are the Maxwell invariants of
electromagnetic fields $F_{\mu \nu }=\partial _{\lbrack \mu }A_{\nu ]}$ and $%
\left( H_{i}\right) _{\mu \nu }=\partial _{\lbrack \mu }\left( B_{i}\right)
_{\nu ]}$, where $A_{\mu }$ and $\left( B_{i}\right) _{\mu }$'s are the
electromagnetic potentials. $\Lambda $, $\lambda _{1}$ and $\lambda _{i}$'s
are constants. The solution for $f(r)$ in Einstein-dilaton gravity governed
by the action (\ref{action}), is \cite{tarrio,Kord}%
\begin{equation}
f(r)=1+\frac{kl^{2}}{z^{2}r^{2}}-\frac{m}{r^{z+2}}+\frac{q^{2}{l}^{2z}{b}%
^{2\left( z-1\right) }}{z{r}^{2\left( z+1\right) }}{,}  \label{Fr}
\end{equation}%
where $m$ and $q$ are two constants which are, respectively, related to
total mass and total charge of the black hole, the dilaton field is%
\begin{equation}
\Phi (r)=\sqrt{z-1}\ln \left( \frac{r}{b}\right) ,  \label{Phi}
\end{equation}%
in which $b$ is a constant and the gauge potentials are%
\begin{gather}
A_{t}=\frac{qb^{2(z-1)}}{z}\left( \frac{1}{r_{+}^{z}}-\frac{1}{r^{z}}\right)
,  \notag \\
\left( B_{2}\right) _{t}=\frac{q_{2}{r}^{z+2}}{\left( z+2\right) {b}^{4}},%
\text{ \ \ \ \ }\left( B_{3}\right) _{t}={\frac{q_{3}{r}^{z}\,}{z{b}^{2}}},
\label{gaupot}
\end{gather}%
where $r_{+}$ is the largest root of metric function $f(r)$, called event
horizon. The constants of the model have been fixed as%
\begin{gather}
\lambda _{1}=-\sqrt{z-1},\text{ \ \ \ \ }\lambda _{2}=\frac{2}{\sqrt{z-1}},%
\text{ \ \ \ \ }\lambda _{3}=\frac{1}{\sqrt{z-1}},  \notag \\
q_{2}^{2}=\frac{\left( z-1\right) (z+2)}{2b^{-4}l^{2z}},\text{ \ \ \ \ }%
q_{3}^{2}=\frac{kb^{2}\left( z-1\right) }{l^{2(z-1)}z},  \notag \\
\Lambda =-\frac{(z+1)(z+2)}{2l^{2}}.  \label{Constants}
\end{gather}%
It is clear from (\ref{Fr}) that $f(r)$ tends to $1$ at spatial infinity and
therefore the metric (\ref{lifmet}) is asymptotically Lifshitz. Looking at $%
q_{3}^{2}$ in (\ref{Constants}), we find that the $k=-1$ case where the
constant curvature at horizon hypersurface is negative, causes an imaginary
charge except for the AdS case with $z=1$ \cite{tarrio,Kord}. Since we will
discuss the Lifshitz solutions with $z>1$, we will consider the cases $k=0$
(black brane) and $k=1$ (black hole) in following studies.

The Hawking temperature can be obtaind as \cite{tarrio,Kord}

\begin{eqnarray}
T =\frac{r_{+}^{z+1}f^{\prime }\left( r_{+}\right) }{4\pi l^{z+1}}&=&\frac{1%
}{4\pi }\left( \frac{\left( z+2\right) m}{l^{z+1}r_{+}^{2}}-\frac{%
2r_{+}^{z-2}k}{z^{2}l^{z-1}}-\frac{2q^{2}\left( z+1\right) }{z{b}^{2\left(
1-z\right) }{l}^{1-z}{r}^{z+2}}\right)  \notag \\
&=&\frac{1}{4\pi }\left( \frac{kr_{+}^{z-2}}{zl^{z-1}}+\frac{\left(
z+2\right) r_{+}^{z}}{l^{z+1}}-\frac{q^{2}{l}^{z-1}{b}^{2\left( z-1\right) }%
}{{r}_{+}^{z+2}}\right) ,
\end{eqnarray}%
where $m$ has been inserted by using the fact that $f(r_{+})=0$. As the
charge of black hole approaches the extreme value%
\begin{equation}
q_{\mathrm{ext}}^{2}=\frac{kr_{+}^{2z}}{zl^{2z-2}{b}^{2\left( z-1\right) }}+%
\frac{\left( z+2\right) {r}_{+}^{2z+2}}{l^{2z}{b}^{2\left( z-1\right) }},
\end{equation}%
the Hawking temperature tends to zero.

In the next section, we will consider a massive charged scalar perturbation
around our Lifshitz solutions and discuss the numerical method to obtain the
corresponding QNM frequencies.

\section{Wave equations for charged scalar perturbation around Lifshitz
solutions}

Here, we intend to consider a massive charged scalar perturbation around our
Lifshitz solutions. The dynamical wave equation for this scalar field
perturbation is 
\begin{equation}
D^{\nu }D_{\nu }\Psi =m_{s}^{2}\Psi ,  \label{ChSc}
\end{equation}%
where $D^{\nu }=\nabla ^{\nu }-iq_{s}A^{\nu }$. The wave function can be
separated into%
\begin{equation}
\Psi =e^{-i\omega t}R(r)Y(\theta ,\phi ),
\end{equation}%
and the differential equation (\ref{ChSc}) can be written into angular part
and radial part by using the metric (\ref{lifmet}) 
\begin{equation}
\nabla ^{2}Y(\theta ,\phi )=-QY(\theta ,\phi ),
\end{equation}%
\begin{equation}
fR^{\prime \prime }+\left( f^{\prime }+\frac{\left( 3+z\right) f}{r}\right)
R^{\prime }+\left( \frac{\omega +q_{s}A_{t}}{r^{z+1}}\right) ^{2}\frac{R}{f}%
-\left( m_{s}^{2}+\frac{Q}{r^{2}}\right) \frac{R}{r^{2}}=0,  \label{Rad}
\end{equation}%
where $Q=\ell (\ell +1)$ and $\ell =0,1,2,\cdots $. Hereafter we will fix
the spacetime radius $l$ to $1$. Defining%
\begin{equation}
R(r)=\frac{K(r)}{r},
\end{equation}%
and the tortoise coordinate 
\begin{equation}
\frac{dr_{\ast }}{dr}=\left[ r^{z+1}f(r)\right] ^{-1},
\end{equation}%
we can rewrite (\ref{Rad}) into the Schr\"{o}dinger form%
\begin{equation}
\frac{d^{2}K(r_{\ast })}{dr_{\ast }^{2}}+\left[ \left( \omega
+q_{s}A_{t}(r)\right) ^{2}-V(r)\right] K(r_{\ast })=0,
\end{equation}%
where the effective potential%
\begin{equation}
V(r)=r^{2z}f(r)\left[ rf^{\prime }(r)+\frac{Q}{r^{2}}+m_{s}^{2}+(z+1)^{2}%
\right] .
\end{equation}%
At spatial infinity $f(r)\rightarrow 1$ and $f^{\prime }(r)\rightarrow 0$,
it is easy to see that $V(r)\rightarrow \infty $ (note that $z\geq 1$) and
therefore we need to impose the Dirichlet boundary condition i.e. $%
R(r)\rightarrow 0$ at the boundary of spatial infinity.

We are going to employ the improved asymptotic iteration method (AIM) \cite%
{ImAIM} to solve (\ref{Rad}) numerically. In order to do so, we rewrite (\ref%
{Rad}) in terms of $u=1-r_{+}/r$ 
\begin{eqnarray}
\frac{r_{+}^{z+1}f(u)}{(1-u)^{z-1}}R^{\prime \prime }(u) &+&\frac{%
r_{+}^{z+1}R^{\prime }(u)}{(1-u)^{z}}\left[ (1-u)f^{\prime }(u)+(z+1)f(u)%
\right]  \notag \\
&+&\frac{R(u)(1-u)^{z-1}}{r_{+}^{z-1}f(u)}\left( q_{s}A_{t}(u)+\omega
\right) ^{2}-\frac{R(u)r_{+}^{z+1}}{(1-u)^{z+1}}\left( m_{s}^{2}+\frac{%
Q(1-u)^{2}}{r_{+}^{2}}\right) =0.  \label{Radu}
\end{eqnarray}%
Considering the asymptotic behaviors of $R(u)$ satisfying (\ref{Radu}), at
horizon ($u=0$) we have $f(0)\approx uf^{\prime }(0)$ and $A_{t}(0)=0 $, so
that (\ref{Radu}) reduces to%
\begin{equation}
R^{\prime \prime }(u)+\frac{R^{\prime }(u)}{u}+\frac{\omega ^{2}R(u)}{%
u^{2}r_{+}^{2z}f^{\prime }(0)^{2}}=0,
\end{equation}%
which has the solution%
\begin{equation}
R(u\rightarrow 0)\sim C_{1}u^{-\xi }+C_{2}u^{\xi },\text{ \ \ \ \ }\xi =%
\frac{i\omega }{r_{+}^{z}f^{\prime }(0)},
\end{equation}%
in which we have to set $C_{2}=0$ to respect the ingoing condition at
horizon.

Near infinity ($u=1$), we have%
\begin{equation}
R^{\prime \prime }(u)+\frac{(z+1)R^{\prime }(u)}{1-u}-\frac{m_{s}^{2}R(u)}{%
(1-u)^{2}}=0,
\end{equation}%
as asymptotic form of (\ref{Radu}) (note that $z\geq 1$). The above
differential equation has the solution

\begin{equation}
R(u\rightarrow 1)\sim D_{1}(1-u)^{\frac{1}{2}\left( z+2+\Pi \right)
}+D_{2}(1-u)^{\frac{1}{2}\left( z+2-\Pi \right) }.
\end{equation}%
where%
\begin{equation}
\Pi =\sqrt{(z+2)^{2}+4m_{s}^{2}}.  \label{Pi}
\end{equation}%
In this case, we should set $D_{2}=0$ in order to satisfy Dirichlet boundary
condition $R(r\rightarrow \infty )\rightarrow 0$.

Now, the desired ansatz for (\ref{Radu}) can be defined as%
\begin{equation}
R(u)=u^{-\xi }(1-u)^{\frac{1}{2}\left( z+2+\Pi \right) }\chi (u).
\label{Ruans}
\end{equation}%
Putting (\ref{Ruans}) into (\ref{Radu}), we have%
\begin{equation}
\chi ^{\prime \prime }=\lambda _{0}(u)\chi ^{\prime }+s_{0}(u)\chi ,
\label{AIM}
\end{equation}%
where%
\begin{equation}
\lambda _{0}(u)=\frac{2i\omega }{ur_{+}^{z}f^{\prime }(0)}-\frac{f^{\prime
}(u)}{f(u)}+\frac{1+\Pi }{1-u},
\end{equation}%
and%
\begin{eqnarray}
s_{0}(u) &=&\frac{r_{+}^{-2(z+1)}f^{\prime }(0)^{-2}}{2(u-1)^{2}u^{2}f(u)^{2}%
}\left[ -2f^{\prime }(0)^{2}r_{+}^{2}u^{2}(1-u)^{2z}(q_{s}A_{t}(u)+\omega
)^{2}\right.  \notag \\
&&\left. +f^{\prime }(0)uf(u)r_{+}^{z}\left( 2f^{\prime }(0)ur_{+}^{z}\left(
m_{s}^{2}r_{+}^{2}+Q(u-1)^{2}\right) -r_{+}^{2}(u-1)f^{\prime }(u)\left(
f^{\prime }(0)u\left( \Pi +z+2\right) r_{+}^{z}-2i(u-1)\omega \right)
\right) \right.  \notag \\
&&\left. +2r_{+}^{2}f(u)^{2}\left( -f^{\prime
}(0)^{2}m_{s}^{2}u^{2}r_{+}^{2z}+if^{\prime }(0)(u-1)\omega r_{+}^{z}\left(
u\Pi +1\right) +(u-1)^{2}\omega ^{2}\right) \right] .
\end{eqnarray}
(\ref{AIM}) can be solved numerically by employing the improved AIM. In the
following, we will set $l=b=1$ and $Q=0$. In \cite%
{LifAIM1,LifAIM2,LifAIM3,LifAIM4}, QNMs corresponding to neutral massive
scalar perturbations around dilaton-Lifshitz solutions have been obtained by
the improved AIM. Here, we will calculate the charged scalar perturbations
and examine the influences of different model parameters on the real and
imaginary parts of quasinormal frequencies around Lifshitz black solutions.

\section{Numerical results \label{nr}}

\begin{figure}[t]
\centering%
\subfigure[~$z=1.2$]{
 \label{fig1a}\includegraphics[width=.31\textwidth]{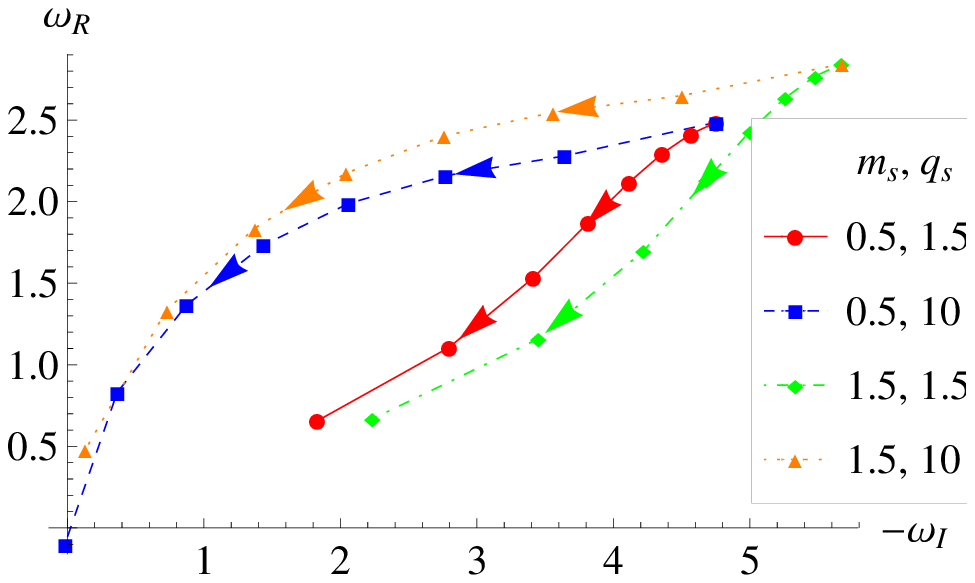} } 
\subfigure[~$z=1.7$]{
 \label{fig1b}\includegraphics[width=.31\textwidth]{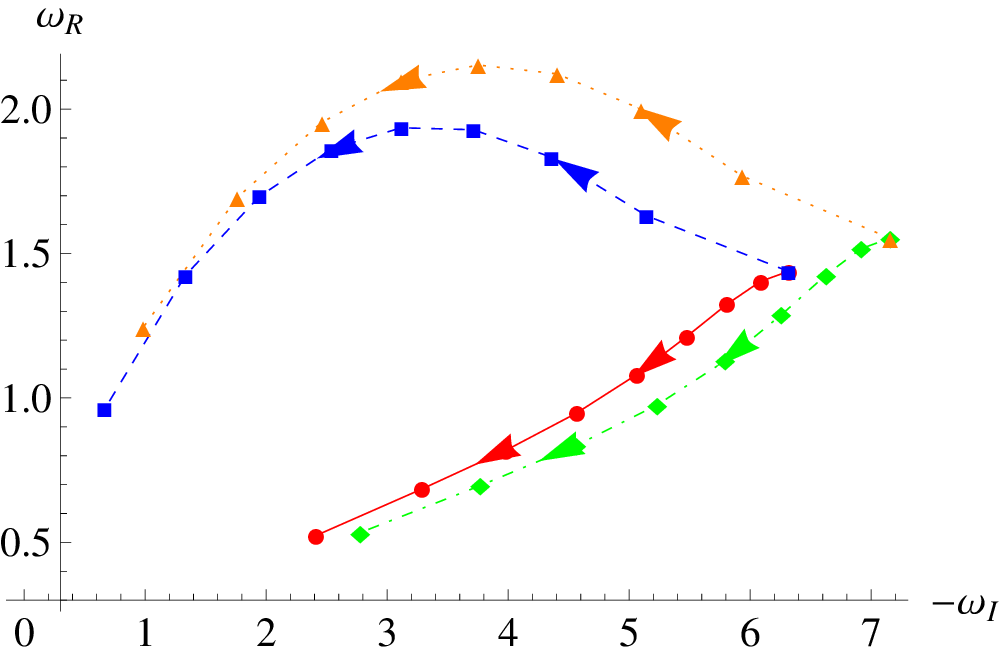} } 
\subfigure[~$z=3$]{
 \label{fig1c}\includegraphics[width=.31\textwidth]{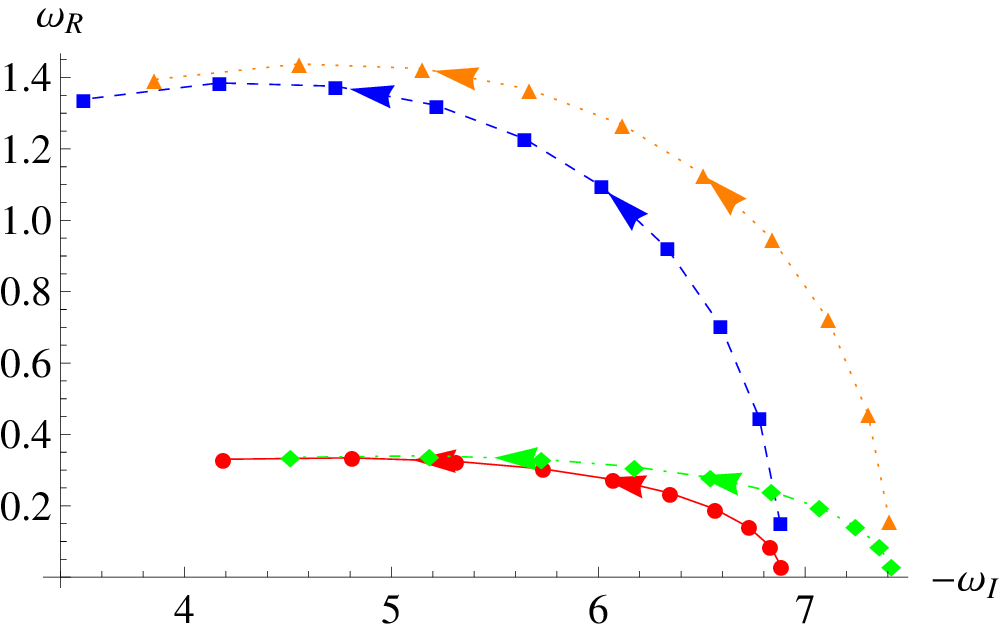} }
\caption{The behaviors of real and imaginary parts of QNFs for different $z$%
's vs $q$ where $k=0$ and $m=3$. The arrows show the direction of the
increase of black branes charge $q$ from $0$ to extreme value $q_{\mathrm{ext%
}}$. The curves with the same colours correspond to same values of $m_{s}$
and $q_{s}$.}
\label{fig1}
\end{figure}

\begin{figure}[t]
\centering%
\subfigure[~$\omega_R$ vs $z$]{
 \label{fig2a}\includegraphics[width=.31\textwidth]{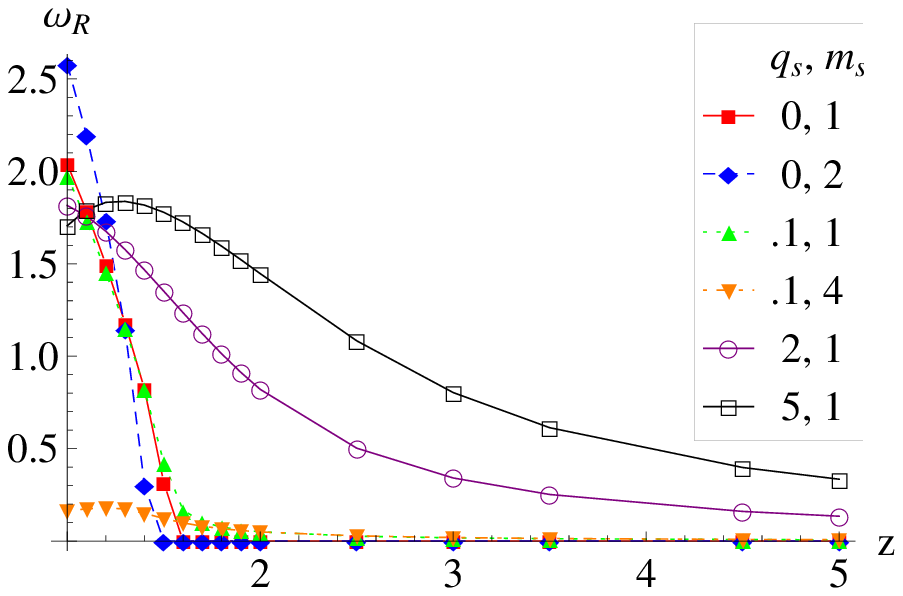} } 
\subfigure[~-$\omega_I$ vs $z$]{
 \label{fig2b}\includegraphics[width=.31\textwidth]{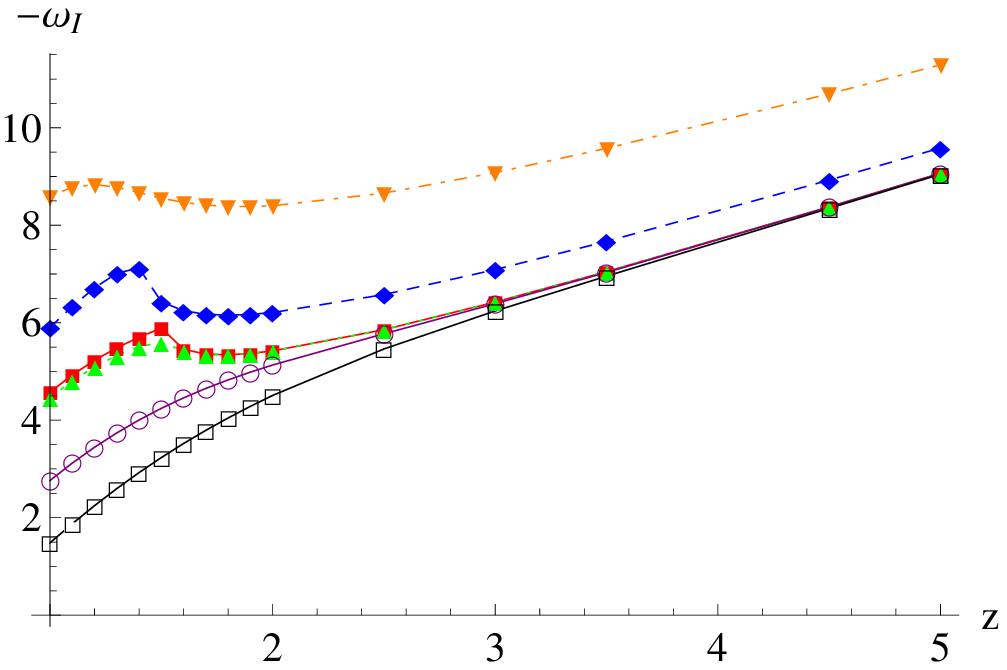} }
\caption{The behaviors of real and imaginary parts of QNMs vs $z$ where $k=0$%
, $m=3$ and $q=1$.}
\label{fig2}
\end{figure}

In this section, we report our numerical results of the QNMs of the charged
scalar perturbation around Lifshitz black brane solutions.

Let us start with the imaginary part of QNMs shown in Fig. \ref{fig1}.
Fixing the mass of the scalar field $m_{s}$ and increasing the charge $q$ of
the background configuration, we observe that the absolute value of the
imaginary part of quasi-normal frequency $|\omega _{I}|$ decreases. This
behavior holds for all chosen values of the charge of the scalar field $q_{s}
$. This property is shown in Fig. \ref{fig1}. Furthermore we find that
fixing the charge of the black hole $q$ and the mass of the scalar field $%
m_{s}$, the bigger value of $q_{s}$ leads to the smaller $|\omega _{I}|$.
But when $q=0$, the $q_{s}$ influence on the imaginary part of the
quasi-normal frequency disappears. The increase of the mass of the scalar
field can enhance $|\omega _{I}|$ when the charges of the scalar field $q_{s}
$ and the background $q$ are fixed.

The decrease of the absolute value of the imaginary part of the quasinormal
frequency $|\omega _{I}|$\ shows that the perturbation outside black hole
will persist longer period of time to decay completely. From holographic
point of view, it means that it takes longer time for the dual system to go
back to equilibrium \cite{30}. When the charge of the black holes is
increased, the $|\omega _{I}|$\ is even smaller which shows that the
perturbation can last even longer to finally vanish outside the black hole.
This result is consistent with the Reissner-Nordstr\"{o}m anti-de Sitter
(RN-AdS) black hole case where it was found in \cite{31} that the absolute
value of the imaginary part of quasinormal frequency $|\omega _{I}|$\
decreases when the black hole charge approaches to the extreme value. In
RN-AdS black hole background when the black hole becomes extreme, $|\omega
_{I}|$\ tends to zero, which makes the extreme RN-AdS black hole marginally
unstable. But in the dilaton-Lifshitz black hole background, the imaginary
part of the quasinormal frequency will not tend to zero except for some big
(small) enough value of the scalar field charge (mass) of the perturbation
field. This exhibits that the compared with the usual RN-AdS black hole, the
stability of the dilaton-Lifshitz black hole is more easily to be protected.
It is important to stress that the black hole stability found here is only
stability against scalar perturbations in a certain region of parameters.

Now, we turn to discuss the behavior of the real part of the quasi-normal
frequency ($\omega _{R}$) exhibited in Fig. \ref{fig1}. When the Lifshitz
exponent $z$ is close to the unity, we find that for fixed $m_{s}$ and $q_{s}
$, $\omega _{R}$ monotonically decreases with the increase of the charge $q$
in the background configuration. Besides, fixing $m_{s}$ and $q$, we observe
that $\omega _{R}$ is smaller when $q_{s}$ becomes bigger, as shown in Fig. %
\ref{fig1a}. It is worth mentioning that the smaller value of the real part
of the quasinormal frequency $\omega _{R}$ shows that the scalar
perturbation have less energy. When $q=0$, the $q_{s}$ influence disappears
as well for the real part of the frequency. When $z$ deviates a bit away
from the unity, for small $q_{s}$, $\omega _{R}$ still decreases with the
increase of black hole charge $q$. But when the scalar field is more charged
with bigger $q_{s}$, $\omega _{R}$ behaves differently and does not
monotonically decrease with the increase of $q$. Instead, there is a barrier
in the value of $\omega _{R}$ so that it increases when $q$ increases from
zero, but then decreases when $q$ is over a critical value (Fig. \ref{fig1b}%
). When $z$ is much bigger, from Fig. \ref{fig1c}, we see that $\omega _{R}$
increases from zero and flattens later with the increase of $q$ from zero
when $q_{s}$ is small. For bigger $q_{s}$, the real part of the frequency
will increase from zero to a maximum value and then slowly decreases when
black hole charge approaches to the extreme value. Comparing with the small $%
z$ case, here for zero black hole charge, we observe a purely damping mode
for charged scalar perturbation. The oscillation adds when $q$ is nonzero.
When the mass of the scalar field is bigger, the difference caused by the
small and large $q_{s}$ will be enlarged in $\omega _{R}$ when the black
hole charge is small. When the black hole charge is big enough, the
influence of the mass of scalar field on $\omega _{R}$ fades away.

In Fig. \ref{fig1a}, we can see that for some choices of parameters, the
imaginary frequency can approach zero and can even jump to be above zero
which shows that the stability of the background configuration can be
destroyed. This phenomenon happens when the charge of the background black
brane solution is high enough, which corresponds to the low temperature. The
instability is consistent with the description of the charged scalar field
condensation to make the original Lifshitz black brane to be a new hairy
configuration. In usual RN-AdS black hole, it was found in \cite{32} that
when the black hole becomes extreme, the imaginary quasinormal frequency
tends to zero, indicating that the extreme RN-AdS black hole is marginally
unstable, since the perturbation will not die out and always persist outside
the black hole background. Then if there comes another stronger wave of
perturbation, the original black hole background is more likely to be
destroyed. This was confirmed in \cite{31}. Here in the dilaton-Lifshitz
black hole background, we observed that when the scalar field is highly
charged, the imaginary quasinormal frequency can even change to be positive.
This shows that instead of the decay of the perturbation outside the
dilaton-Lifshitz black hole background, the highly charged scalar
perturbation can even blow up outside the black hole and destroy the
original black hole spacetime structure. This effect to make the spacetime
unstable brought in by the highly charged perturbation field was also
observed in the stability analysis in Reissner-Nordstr\"{o}m black hole in
de Sitter background \cite{1405.4931}.

As one can see from Fig. \ref{fig1a}, when the Lifshitz exponent $z$ is
close to unity, and decreasing $q_{s}$ or increasing $m_{s}$ can keep the
imaginary frequency to be negative to ensure the stability. When $z$ is more
away from unity (Figs. \ref{fig1b} and \ref{fig1a}) for given values of $%
m_{s}$ and $q_{s}$, there is no possibility to destroy the original
background configuration, since even when $q$ takes the extreme value, the
imaginary frequency is negative. For chosen value of the Lifshitz exponent $z
$, we can find threshold values for the mass and charge of scalar
perturbation field to ensure the stability of the original Lifshitz black
brane when its charge $q$ is nearly extreme. These threshold values are
listed in table \ref{tab1}. For fixing $m_{s}$, when $q_{s}$ is below the
corresponding threshold values for the chosen $z$, the black brane can keep
to be stable. When we fix $q_{s}$, the mass of the scalar field is above the
corresponding threshold value for the chosen $z$ can ensure the stability. $%
m_{s}^{2}$ has a lower bound $-(z+2)^{2}/4$ according to Eq. (\ref{Pi}).
Thus $m_{s}$ cannot be reduced infinitely so that in table \ref{tab1} at
some $z$ the decrease of $m_{s}^{2}$ stops.

\begin{table}[h]
\caption{The threshold values of $m_{s}$ and $q_{s}$ for chosen $z$ to keep
the stability of the black brane with extreme charge value $q$ and $m=3$. }
\label{tab1}
\begin{center}
\begin{tabular}{|c|c|c|c|c|c|c|c|c|c|c|c|}
\hline
\multicolumn{6}{|l}{$m_{s}=0.6$} & \multicolumn{6}{|l|}{$q_{s}=8.8$} \\ 
\hline
$z$ & $1$ & $1.2$ & $1.6$ & $2$ & $3$ & $z$ & $1$ & $1.2$ & $1.6$ & $1.7$ & $%
1.8$ \\ \hline
$q_{s}$ & $5.3$ & $7.3$ & $8.8$ & $12.3$ & $22$ & $m_{s}^{2}$ & $6.25$ & $%
2.25$ & $0.36$ & $-3.42$ & $-$ \\ \hline
\end{tabular}%
\end{center}
\end{table}

It is clear from above discussions that increasing $z$\ makes the stability
more easily to be protected. This result is physically reasonable from
holographic point of view since it is in agreement with the behavior of
Lifshitz superconductors where it was found that increasing dynamical
critical exponent $z$\ makes superconductor more difficult to be formed \cite%
{lifsup}.

In Fig. \ref{fig2}, the behaviors of real and imaginary parts of
quasi-normal frequencies with the change of Lifshitz exponent $z$ have been
illustrated. In \cite{LifAIM1}, it was shown that for ($3+1$)-dimensional
uncharged dilaton-Lifshitz black branes, independent of the value of $m_{s}$%
, QNMs corresponding to neutral scalar perturbations start overdamping (with 
$\omega _{R}=0$) at $z=2$. Here, we find that for charged black branes, the
point at which the real part of quasi-normal frequency of neutral scalar
perturbation starts to vanish is no longer at $z=2$, but depends on the
value of $m_{s}$ (see red and blue lines in Fig. \ref{fig2a}). For fixed $%
m_{s}$, when $q_{s}$ is small, $\omega _{R}$ reduces and finally flattens
when the Lifshitz exponent $z$ becomes big. The value of $\omega _{R}$ for
high $z$ can approach to zero (see red and green lines in Fig. \ref{fig2a}).
For bigger $q_{s}$, $\omega _{R}$ finally will be above zero (see purple
line in Fig. \ref{fig2a}).

Now, we discuss the imaginary part of quasi-normal frequency as shown in
Fig. \ref{fig2}. For small $q_{s}$, the absolute value of the imaginary part
is always bigger than the corresponding value with bigger $q_s$. With the
increase of the Lifshitz exponent $z$, the absolute value of $\omega _{I}$
increases. For large $z$, $|\omega _{I}|$ has little dependence on $q_{s}$
and it is mainly influenced by the value of $m_{s}$.

We have mainly reported QNM behaviors of the charged scalar perturbation in
the Lifshitz black brane background with $k=0$. For the Lifshitz black hole
case with $k=1$ we have found similar results. We do not repeat explaining
these similar results here for Lifshitz black holes.

\section{Summary and conclusion}

One of the important subject in black hole physics is to investigate the
resonances for the scattering of incoming waves by black holes. Quasi-normal
modes (QNMs) of a black hole spacetime are indeed the proper solutions of
the perturbation equations. It is a general belief that QNMs carry unique
footprints to directly identify the black hole existence.

Most available studies on QNMs corresponding to Lifshitz solutions are
restricted to neutral scalar perturbations. In this letter, we have extended
the study to the charged scalar perturbation in the background of ($3+1$%
)-dimensional charged dilaton Lifshitz branes/holes. Asymptotic Lifshitz
solutions are interesting duals to many condensed matter systems.

We have used the improved asymptotic iteration method (AIM) to calculate the
charged scalar perturbations numerically and examine the influences of
different model parameters on the imaginary and real parts of quasi-normal
frequencies of the charged scalar field perturbations around Lifshitz black
solutions. In contrast to the case of neutral scalar perturbations, we found
that it is possible to destroy the Lifshitz configuration. To be more clear,
we observed that for suitable choices of model parameters, imaginary part of
QNM frequencies can be positive near extreme charge value of Lifshitz
configuration where temperature is low. For a chosen $z$, there are
threshold values for mass and charge of scalar perturbation to guarantee the
stability of the Lifshitz black configuration. Fixing $q_{s}$ and taking the
mass of the scalar field above the corresponding threshold value, or fixing $%
m_{s}$ and keeping the charge of the scalar field below the threshold value,
the black brane/hole stability can be protected. It is remarkable to note
that here by stability of black hole, we mean only the stability against
scalar perturbations in a certain region of parameters. In terms of $z$, it
was shown that when it is more away from unity, it is more difficult for the
Lifshitz background to become unstable. This result is consistent with the
observation that Lifshitz superconductors are more difficult to be formed
for greater values of Lifshitz exponent $z$ \cite{lifsup}. We have also
observed rich dependence of the quasinormal fequencies on the mass of the
scalar field $m_{s}$, the charge $q$ of the background configuration and the
charge of the scalar field $q_{s}$.

We have also investigated the behaviors of the real and imaginary parts of
quasi-normal frequencies in terms of the Lifshitz exponent $z$. We found out
that for charged black branes, the point at which neutral QNMs start
overdamping is no longer at $z=2$ as what was claimed in \cite{LifAIM1} for
uncharged Lifshitz black branes, but it has dependence on the value of $%
m_{s} $. For fixed $m_{s}$, when $q_{s} $ is small, $\omega _{R}$ reduces
and finally flattens when the Lifshitz exponent $z$ becomes big. The value
of $\omega _{R}$ for high $z$ can finally approach to zero, whereas for
bigger $q_{s}$, this possibility does not exist and $\omega _{R}$ will be
finally above zero. The rich dependence of the imaginary part $|\omega _{I}|$
on the Lifshitz exponent $z$, the scalar field charge $q_{s}$, mass $m_{s}$
has also been investigated carefully. We have found that the QNMs properties
of the Lifshitz black brane background with $k=0$ hold as well for the
Lifshitz black hole case with $k=1$. To be concise, we do not repeat the
discussion for the Lifshitz black hole here.

Finally, it is worth mentioning that in this letter we have only considered
the gauge field as the linear Maxwell field. It is interesting to generalize
this study to other nonlinear gauge fields such as power-law Maxwell,
Born-Infeld, exponential and logarithmic nonlinear gauge fields in the
background of Lifshitz spacetime. Besides, we studied the $(3+1)$%
-dimensional charged dilaton Lifshitz black solutions. It is also
interesting if one could extend the study to higher dimensional Lifshitz
spacetime. In addition, we investigated the scalar perturbations. It is
worthwhile to extend this investigation to the vector and tensor
perturbations in this context. These issues are now under investigation and
the results will appear in our future works.

\begin{acknowledgments}
We thank the anonymous referee for constructive comments which helped us
improve the letter significantly. MKZ is grateful to C. Y. Zhang for helpful
discussions. MKZ and AS thank the research council of Shiraz University. The
work of BW was partially supported by NNSF of China. This work has been
supported financially by Research Institute for Astronomy \& Astrophysics of
Maragha (RIAAM).
\end{acknowledgments}


\begin{thebibliography}{99}
\bibitem{qnm4} K. D. Kokkotas and B. G. Schmidt, \textit{Quasi-normal modes
of stars and black holes}, Living Rev. Rel. \textbf{2}, 2 (1999)
[gr-qc/9909058].

\bibitem{qnm5} H. P. Nollert, \textit{Quasinormal modes: the characteristic
sound of black holes and neutron stars}, Class. Quant. Grav. \textbf{16},
R159 (1999).

\bibitem{qnm6} R. A. Konoplya and A. Zhidenko, \textit{Quasinormal modes of
black holes: from astrophysics to string theory}, Rev. Mod. Phys. \textbf{83}
,793 (2011) [arXiv:1102.4014].

\bibitem{wang1} B. Wang, \textit{Perturbations around black holes}, Braz. J.
Phys. \textbf{35}, 1029 (2005) [gr-qc/0511133].

\bibitem{3} B. Wang, C. Y. Lin and E. Abdalla, \textit{Quasinormal modes of
Reissner-Nordstr\"{o}m anti-de Sitter black holes}, Phys. Lett. B \textbf{481%
}, 79 (2000) [hep-th/0003295];\newline
B. Wang, C. Molina and E. Abdalla, \textit{Evolution of a massless scalar
field in Reissner-Nordstr\"{o}m anti-de Sitter spacetimes}, Phys. Rev. D 
\textbf{63}, 084001 (2001) [hep-th/0005143];\newline
J. M. Zhu, B.Wang and E. Abdalla, \textit{Object picture of quasinormal
ringing on the background of small Schwarzschild anti-de Sitter black holes}%
, Phys. Rev. D \textbf{63}, 124004 (2001) [hep-th/0101133];\newline
V. Cardoso and J. P. S. Lemos, \textit{Scalar, electromagnetic, and Weyl
perturbations of BTZ black holes: Quasinormal modes}, Phys. Rev. D \textbf{63%
}, 124015 (2001) [gr-qc/0101052];\newline
V. Cardoso and J. P. S. Lemos, \textit{Quasinormal modes of
Schwarzschild-anti-de Sitter black holes: Electromagnetic and gravitational
perturbations}, Phys. Rev. D \textbf{64}, 084017 (2001) [gr-qc/0105103];%
\newline
V. Cardoso and J. P. S. Lemos, \textit{Quasi-normal modes of toroidal,
cylindrical and planar black holes in anti-de Sitter spacetimes: scalar,
electromagnetic and gravitational perturbations}, Class. Quantum Grav.%
\textbf{\ 18}, 5257 (2001) [gr-qc/0107098];\newline
E. Winstanley, \textit{Classical super-radiance in Kerr-Newman-anti-de
Sitter black holes}, Phys. Rev. D \textbf{64}, 104010 (2001) [gr-qc/0106032];%
\newline
J. Crisstomo, S. Lepe and J. Saavedra, \textit{Quasinormal modes of the
extremal BTZ black hole}, Class. Quant. Grav. \textbf{21}, 2801 (2004)
[hep-th/0402048];\newline
S. Lepe, F. Mendez, J. Saavedra and L. Vergara, \textit{Fermions scattering
in a three-dimensional extreme black-hole background}, Class. Quant. Grav. 
\textbf{20}, 2417 (2003) [hep-th/0302035];\newline
D. Birmingham, I. Sachs and S. N. Solodukhin, \textit{Conformal field theory
interpretation of black hole quasinormal modes}, Phys. Rev. Lett. \textbf{88}%
, 151301 (2002) [hep-th/0112055];\newline
D. Birmingham, \textit{Choptuik scaling and quasinormal modes in the anti-de
Sitter space/conformal-field theory correspondence}, Phys. Rev. D \textbf{64}%
, 064024 (2001) [hep-th/0101194];\newline
B. Wang, E. Abdalla and R. B. Mann, \textit{Scalar wave propagation in
topological black hole backgrounds}, Phys. Rev. D \textbf{65}, 084006 (2002)
[hep-th/0107243];\newline
J. S. F. Chan and R. B. Mann, \textit{Scalar wave falloff in topological
black hole backgrounds}, Phys. Rev. D \textbf{59}, 064025 (1999);\newline
S. Musiri and G. Siopsis, \textit{Asymptotic form of quasi-normal modes of
large AdS black holes}, Phys. Lett. B \textbf{576}, 309 (2003)
[hep-th/0308196];\newline
R. Aros, C. Martinez, R. Troncoso and J. Zanelli, \textit{Quasinormal modes
for massless topological black holes}, Phys. Rev. D \textbf{67}, 044014
(2003) [hep-th/0211024];\newline
A. Nunez, A. O. Starinets, \textit{AdS/CFT correspondence, quasinormal
modes, and thermal correlators in }$N=4$\textit{\ supersymmetric Yang-Mills
theory}, Phys. Rev. D \textbf{67}, 124013 (2003) [hep-th/0302026].

\bibitem{30} G. T. Horowitz and V. E. Hubeny, \textit{Quasinormal Modes of
AdS Black Holes and the Approach to Thermal Equilibrium}, Phys. Rev. D 
\textbf{62}, 024027 (2000) [hep-th/9909056].

\bibitem{31} B. Wang, C. Y. Lin and C. Molina, \textit{Quasinormal behavior
of massless scalar field perturbation in Reissner-Nordstr\"{o}m anti-de
Sitter spacetimes}, Phys. Rev. D \textbf{70}, 064025 (2004) [hep-th/0407024].

\bibitem{32} E. Berti and K. D. Kokkotas, \textit{Quasinormal modes of
Reissner-Nordstr\"{o}m-anti-de Sitter black holes: Scalar, electromagnetic,
and gravitational perturbations}, Phys. Rev. D \textbf{67}, 064020 (2003)
[gr-qc/0301052].

\bibitem{4} E. Abdalla, B. Wang, A. Lima-Santos and W. G. Qiu, \textit{%
Support of dS/CFT correspondence from perturbations of three-dimensional
spacetime}, Phys. Lett. B \textbf{538}, 435 (2002) [hep-th/0204030];\newline
E. Abdalla, K. H. CastelloBranco and A. Lima-Santos, \textit{Support of
dS/CFT correspondence from space-time perturbations}, Phys. Rev. D \textbf{66%
}, 104018 (2002) [hep-th/0208065].

\bibitem{5} S. Hod, \textit{Bohr's correspondence principle and the area
spectrum of quantum black holes}, Phys. Rev. Lett.\textbf{\ 81}, 4293 (1998)
[gr-qc/9812002];\newline
A. Corichi, \textit{Quasinormal modes, black hole entropy, and quantum
geometry}, Phys. Rev. D \textbf{67}, 087502 (2003) [gr-qc/0212126];\newline
L. Motl, \textit{An analytical computation of asymptotic Schwarzschild
quasinormal frequencies}, Adv. Theor. Math. Phys. \textbf{6}, 1135 (2003)
[gr-qc/0212096];\newline
L. Motl and A. Neitzke, \textit{Asymptotic black hole quasinormal frequencies%
}, Adv. Theor. Math. Phys. \textbf{7}, 307 (2003) [hep-th/0301173];\newline
A. Maassen van den Brink, \textit{WKB analysis of the Regge-Wheeler equation
down in the frequency plane}, J. Math. Phys. \textbf{45}, 327 (2004)
[gr-qc/0303095];\newline
O. Dreyer, \textit{Quasinormal modes, the area spectrum, and black hole
entropy}, Phys. Rev. Lett. \textbf{\ 90}, 08130 (2003) [gr-qc/0211076];%
\newline
G. Kunstatter, $d$\textit{-dimensional black hole entropy spectrum from
quasinormal modes}, Phys. Rev. Lett. \textbf{90}, 161301 (2003)
[gr-qc/0212014];\newline
N. Andersson and C. J. Howls, \textit{The asymptotic quasinormal mode
spectrum of non-rotating black holes}, Class. Quant. Grav. \textbf{21}, 1623
(2004) [gr-qc/0307020];\newline
V. Cardoso, J. Natario and R. Schiappa, \textit{Asymptotic quasinormal
frequencies for black holes in nonasymptotically flat space-times}, J. Math.
Phys. \textbf{45}, 4698 (2004) [hep-th/0403132];\newline
J. Natario and R. Schiappa, \textit{On the classification of asymptotic
quasinormal frequencies for }$d$\textit{-dimensional black holes and quantum
gravity,} Adv. Theor. Math. Phys. \textbf{8}, 1001 (2004) [hep-th/0411267];%
\newline
V. Cardoso and J. P. S. Lemos, \textit{Quasinormal modes of the near
extremal Schwarzschild-de Sitter black hole}, Phys. Rev. D \textbf{67},
084020 (2003) [gr-qc/0301078];\newline
K. H. C. CastelloBranco and E. Abdalla, \textit{Analytic determination of
the asymptotic quasi-normal mode spectrum of small Schwarzschild-de Sitter
black holes}, gr-qc/0309090.

\bibitem{6} G. Koutsoumbas, S. Musiri, E. Papantonopoulos and G. Siopsis, 
\textit{Quasi-normal modes of electromagnetic perturbations of
four-dimensional topological black holes with scalar hair}, JHEP \textbf{0610%
}, 006 (2006) [hep-th/0606096];\newline
X. P. Rao, B. Wang and G. H. Yang, \textit{Quasinormal modes and phase
Transition of black holes}, Phys. Lett. B \textbf{649}, 472 (2007)
[arXiv:0712.0645];\newline
R. G. Cai, Z. Y. Nie, B. Wang and H. Q. Zhang, \textit{Quasinormal Modes of
Charged Fermions and Phase Transition of Black Holes}, arXiv:1005.1233;%
\newline
Y. Liu, D. C. Zou and B. Wang, \textit{Signature of the Van der Waals like
small-large charged AdS black hole phase transition in quasinormal modes},
JHEP \textbf{1409}, 179 (2014) [arXiv:1405.2644];\newline
J. Shen, B. Wang, C. Y. Lin, R. G. Cai and R. K. Su, \textit{The phase
transition and the Quasi-Normal Modes of black Holes}, JHEP \textbf{0707},
037 (2007) [hep-th/0703102].

\bibitem{Lif} S. Kachru, X. Liu and M. Mulligan, \textit{Gravity Duals of
Lifshitz-like Fixed Points}, Phys. Rev. D \textbf{78}, 106005 (2008)
[arXiv:0808.1725].

\bibitem{hcc1} E. Ay\'{o}n-Beato, A. Garbarz, G. Giribet and M. Hassa\"{\i}%
ne, \textit{Lifshitz black hole in three dimensions}, Phys. Rev. D \textbf{80%
}, 104029 (2009) [arXiv:0909.1347].

\bibitem{hcc2} E. Ay\'{o}n-Beato, A. Garbarz, G. Giribet and M. Hassa\"{\i}%
ne, \textit{Analytic Lifshitz black holes in higher dimensions}, JHEP 
\textbf{1004}, 030 (2010) [arXiv:1001.2361].

\bibitem{hcc3} H. Lu, Y. Pang, C. N. Pope and J. F. Vazquez-Poritz, \textit{%
AdS and Lifshitz black holes in conformal and Einstein-Weyl gravities},
Phys. Rev. D \textbf{86}, 044011 (2012) [arXiv:1204.1062].

\bibitem{massg1} K. Balasubramanian and J. McGreevy, \textit{An Analytic
Lifshitz black hole}, Phys. Rev. D \textbf{80}, 104039 (2009)
[arXiv:0909.0263].

\bibitem{tarrio} J. Tarrio and S. Vandoren, \textit{Black holes and black
branes in Lifshitz spacetimes}, JHEP \textbf{1109, }017 (2011)
[arXiv:1105.6335].

\bibitem{Kord} M. Kord Zangeneh, A. Sheykhi and M. H. Dehghani, \textit{%
Thermodynamics of topological nonlinear charged Lifshitz black holes, }Phys.
Rev. D\ \textbf{92}, 024050 (2015) [arXiv:1506.01784].

\bibitem{Kord1} M. Kord Zangeneh, M. H. Dehghani and A. Sheykhi, \textit{%
Thermodynamics of Gauss-Bonnet-Dilaton Lifshitz Black Branes, }Phys. Rev. D%
\textbf{\ 92,} 064023 (2015) [arXiv:1506.07068].

\bibitem{Kord2} M. Kord Zangeneh, A. Dehyadegari, A. Sheykhi and M. H.
Dehghani, \textit{Thermodynamics and gauge/gravity duality for Lifshitz
black holes in the presence of exponential electrodynamics, }JHEP \textbf{%
1603,} 037 (2016) [arXiv:1601.04732].

\bibitem{Kord3} A. Dehyadegaria, A. Sheykhi and M. Kord Zangeneh, \textit{%
Holographic conductivity for logarithmic charged dilaton-Lifshitz solutions, 
}Phys. Lett. B\textbf{\ 758,} 226 (2016) [arXiv:1602.08476].

\bibitem{Kord4} M. Kord Zangeneh, A. Dehyadegari, M. R. Mehdizadeh, B. Wang
and A. Sheykhi, \textit{Thermodynamics, phase transitions and Ruppeiner
geometry for Einstein-dilaton Lifshitz black holes in the presence of
Maxwell and Born-Infeld electrodynamics}, arXiv:1610.06352.

\bibitem{nonAb} X. H. Feng and W. J. Geng, \textit{Non-abelian (hyperscaling
violating) Lifshitz black holes in general dimensions}, Phys. Lett. B 
\textbf{747}, 395 (2015) [arXiv:1502.00863].

\bibitem{nmg1} B. Cuadros-Melgar, J. de Oliveira and C. E. Pellicer, \textit{%
Stability Analysis and Area Spectrum of }$3$\textit{-Dimensional Lifshitz
Black Holes}, Phys. Rev. D \textbf{85}, 024014 (2012) [arXiv:1110.4856].

\bibitem{nmg2} B. Cuadros-Melgar, J. de Oliveira and C. E. Pellicer, \textit{%
Quasinormal modes and thermodynamical aspects of the }$3$\textit{D Lifshitz
black hole}, J. Phys. Conf. Ser. \textbf{453}, 012025 (2013)
[arXiv:1302.6185].

\bibitem{hcc4} A. Giacomini, G. Giribet, M. Leston, J. Oliva and S. Ray, 
\textit{Scalar field perturbations in asymptotically Lifshitz black holes},
Phys. Rev. D \textbf{85}, 124001 (2012) [arXiv:1203.0582].

\bibitem{hcc5} E. Abdalla, O. P. F. Piedra, F. S. Nu\~{n}ez and J. de
Oliveira, \textit{Scalar field propagation in higher dimensional black holes
at a Lifshitz point}, Phys. Rev. D \textbf{88}, 064035 (2013)
[arXiv:1211.3390].

\bibitem{hcc6} M. Catalan, E. Cisternas, P. A. Gonzalez and Y. Vasquez, 
\textit{Quasinormal modes and greybody factors of a four-dimensional
Lifshitz black hole with }$z=0$, Astrophysics and Space Sci. \textbf{361}, 1
(2016) [arXiv:1404.3172].

\bibitem{pro} P. A. Gonzalez, J. Saavedra and Y. Vasquez, \textit{%
Quasinormal modes and Stability Analysis for Four-dimensional Lifshitz Black
Hole}, International J. of Modern Phys. D \textbf{21}, 1250054 (2012)
[arXiv:1201.4521].

\bibitem{Mann} R. B. Mann, \textit{Lifshitz topological black holes}, JHEP 
\textbf{0906}, 075 (2009) [arXiv:0905.1136].

\bibitem{topqnm} P. A. Gonzalez, F. Moncada and Y. Vasquez, \textit{%
Quasinormal Modes, Stability Analysis and Absorption Cross Section for }$4$%
\textit{-dimensional Topological Lifshitz Black Hole}, Euro. Phys. J. C 
\textbf{72}, 1 (2012) [arXiv:1205.0582].

\bibitem{myung} Y. S. Myung and T. Moon, \textit{Quasinormal frequencies and
thermodynamic quantities for the Lifshitz black holes}, Phys. Rev. D \textbf{%
86}, 024006 (2012) [arXiv:1204.2116].

\bibitem{LifAIM1} W. Sybesma and S. Vandoren, \textit{Lifshitz quasinormal
modes and relaxation from holography}, JHEP \textbf{1505}, 021 (2015)
[arXiv:1503.07457].

\bibitem{LifAIM2} P. A. Gonz\'{a}lez and Y. V\'{a}squez, \textit{Scalar
Perturbations of Nonlinear Charged Lifshitz Black Branes with Hyperscaling
Violation}, Space Sci. \textbf{361}, 224 (2016) [arXiv:1509.00802].

\bibitem{LifAIM4} R. Becar, P. A. Gonz\'{a}lez and Y. V\'{a}squez, \textit{%
Quasinormal modes of non-Abelian hyperscaling violating Lifshitz black holes}%
, arXiv:1510.04605.

\bibitem{LifAIM3} R. Becar, P. A. Gonz\'{a}lez and Y. V\'{a}squez, \textit{%
Quasinormal modes of four-dimensional topological nonlinear charged Lifshitz
black holes}, Eur. Phys. J. C \textbf{76}, 78 (2016) [arXiv:1510.06012].

\bibitem{park} C. Park, \textit{A Massive Quasi-normal Mode in the
Holographic Lifshitz Theory}, Phys. Rev. D \textbf{89}, 066003 (2014)
[arXiv:1312.0826].

\bibitem{ferm} M. Catalan, E. Cisternas, P. A. Gonzalez and Y. Vasquez, 
\textit{Dirac quasinormal modes for a }$4$\textit{-dimensional Lifshitz
Black Hole}, Euro. Phys. J. C \textbf{74}, 1 (2014) [arXiv:1312.6451].

\bibitem{elec} A. Lopez-Ortega, \textit{Electromagnetic quasinormal modes of
an asymptotically Lifshitz black hole}, Gen. Rel. and Grav. \textbf{46}, 1
(2014) [arXiv:1406.0126].

\bibitem{konop} R. A. Konoplya, \textit{Decay of a charged scalar field
around a black hole: Quasinormal modes of RN, RNAdS, and dilaton black holes}%
, Phys. Rev. D \textbf{66}, 084007 (2002) [gr-qc/0207028]; \newline
R. A. Konoplya and A. Zhidenko, \textit{Decay of a charged scalar and Dirac
fields in the Kerr-Newman-de Sitter background}, Phys. Rev. D \textbf{76},
084018 (2007) [Erratum: Phys. Rev. D \textbf{90}, 029901 (2014)]
[arXiv:0707.1890]; \newline
R. A. Konoplya and A. Zhidenko, \textit{Massive charged scalar field in the
Kerr-Newman background I: quasinormal modes, latetime tails and stability},
Phys. Rev. D \textbf{88}, 024054 (2013) [arXiv:1307.1812].

\bibitem{1405.4931} Z. Zhu, S. J. Zhang, C. E. Pellicer, B. Wang and E.
Abdalla, \textit{Stability of Reissner-Nordstr\"{o}m black hole in de Sitter
background under charged scalar perturbation}, Phys. Rev. D \textbf{90},
044042 (2014) [arXiv:1405.4931].

\bibitem{konop1} R. A. Konoplya and A. Zhidenko, \textit{Charged scalar
field instability between the event and cosmological horizons}, Phys. Rev. D 
\textbf{90}, 064048 (2014) [arXiv:1406.0019].

\bibitem{1002.2679} X. He, B. Wang, R. G. Cai and C. Y. Lin, \textit{%
Signature of the black hole phase transition in quasinormal modes}, Phys.
Lett. B \textbf{688}, 230 (2010) [arXiv:1002.2679].

\bibitem{1010.2806} E. Abdalla, C. E. Pellicer, J. de Oliveira and A. B.
Pavan, \textit{Phase transitions and regions of stability in Reissner-Nordstr%
\"{o}m holographic superconductors}, Phys. Rev. D \textbf{82}, 124033 (2010)
[arXiv:1010.2806].

\bibitem{1111.6729} Y. Liu and B. Wang, \textit{Perturbations around the AdS
Born-Infeld black holes}, Phys. Rev. D \textbf{85}, 046011 (2012)
[arXiv:1111.6729].

\bibitem{ImAIM} H. T. Cho, A. S. Cornell, J. Doukas and W. Naylor, \textit{%
Black hole quasinormal modes using the asymptotic iteration method}, Class.
Quant. Grav. \textbf{27}, 155004 (2010) [arXiv:0912.2740].

\bibitem{lifsup} J. W. Lu, Y. B. Wu, P. Qian, Y. Y. Zhao, X. Zhang and N.
Zhang, \textit{Lifshitz Scaling Effects on Holographic Superconductors},
Nucl. Phys. B \textbf{887}, 112 (2014) [arXiv:1311.2699].
\end{thebibliography}
\end{document}